\newcommand{\act}{\omega}           
\newcommand{\CLT}{$\mathcal{T}$}   
\newcommand{\az}{z\,\act_{\infty}}  
\newcommand{\CLO}{$\mathcal{O}$}   
\newcommand{\oned}{one dimensional }
\newcommand{\tia}{t_{\rm ia}}       
\newcommand{\tev}{\tau}    
\newcommand{\tav}{\tau^{\rm av}}    

\documentstyle[12pt,psfig]{iopart}  

\begin{document}

\baselineskip 14pt

\jl{1}

\title{Zero-temperature criticality in a simple glass model}

\author{David Head\footnote[2]{Electronic address: david@ph.ed.ac.uk}}

\address{Department of Physics and Astronomy, University of Edinburgh,
JCMB King's Buildings, Mayfield Road, Edinburgh EH9 3JZ, UNITED KINGDOM.}
\begin{abstract}
We introduce the strongly-interacting trap model, a
version of Bouchaud's trap model for glasses [Bouchaud J-P 1992
{\em J. Phys. I France {\bf 2}} 1705].
At finite temperatures the model exhibits glassy relaxation over
intermediate timeframes but reaches a steady state at finite times.
In limit of zero temperature and with a suitably renormalised
timescale the model maps onto the Bak-Sneppen model,
widely studied in the context of self-organised criticality
[Bak P and Sneppen K 1993 {\em Phys. Rev. Lett. {\bf 71}} 4083].
Hence zero temperature is a critical point in all dimensions.
These claims are supported by mean field analysis of the stationary
solution and numerical simulations of a finite-dimensional lattice model.
\end{abstract}
\maketitle



\section{Introduction}

Our understanding of complex physical systems is often aided by
the construction of simple mathematical models.
A prime example from glass theory is the trap model of
Bouchaud~\cite{bouchaud,monthus_and_bouchaud}, which
describes a system's relaxation purely in terms of
activated processes between potential energy wells of various depths.
With only a few assumptions the model manages to
reproduce many of the phenomena commonly
associated with glasses, such as a crossover to non-equilibrium
behaviour as the temperature is lowered and aging
in the low-temperature regime.
Furthermore the model is sufficiently generic to apply to
structural as well as spin glasses, and has recently been extended to
the study of rheology in glassy systems~\cite{rheo1}-\cite{rheo4}.

However, there are a number of spin-glass models that exhibit a continuous
phase transition, or critical point, in the limit of zero
temperature, quite unlike the current realisation of Bouchaud's trap model.
These include Stein and Newman's model, which
maps onto invasion percolation at a
temperature \mbox{$T=0$}~\cite{stein1,stein2};
the two dimensional Ising spin glass~\cite{ising_sg};
and the Sherrington-Kirkpatrick model under infinitesimal
variations of the external magnetic field~\cite{pazmandi}.

In this paper we introduce a modified version of Bouchaud's trap model
which incorporates a form of interaction between different subsystems.
It is shown that in the limit of zero temperature this model maps
onto a self-organised critical system known as the Bak-Sneppen
model~\cite{BS,BSRev}, so \mbox{$T=0^{+}$} is a critical point in
all dimensions, including the infinite-dimensional mean field model.
Although the Bak-Sneppen model was originally devised to explain the
pattern of evolutionary bursts in the fossil record known
as `punctuated equilibria,'
it was justified in terms of activated processes over (fitness)
barriers and so its biological interpretation is purely semantical.

This paper is arranged as follows.
In section \ref{s:mf_analysis}, Bouchaud's model is briefly
reviewed before the new model is defined in its mean field form.
The corresponding master equation is solved in the steady state.
The relationship with the Bak-Sneppen model implies that the model
can also be defined on a lattice, and this modification is studied
numerically in section \ref{s:lattice}.
The simulations were also employed to check the analytical predictions
of the mean field equations.
Finally, our work is summarised in section \ref{s:conc}.

\section{Definition of the mean field model and the stationary solution}
\label{s:mf_analysis}

In Bouchaud's formalism, the state of the system is represented by a particle
(or many particles, if the system is treated as a collection of subsystems)
moving in configuration space over an energy landscape,
which is `rugged' in the sense that most of the particle's time
is spent trapped in potential wells or
`traps'~\cite{bouchaud,monthus_and_bouchaud}.
Motion within traps is not explicitly incorporated;
instead, it is simply assumed that whatever dynamics is present
gives rise an activation probability that depends on the
barrier height $b$ and the temperature $T$
according to the Arrhenius form \mbox{$\act_{0}\,{\e}^{-b/T}$},
where $\act_{0}$ fixes the timescale.

When a particle is activated it `hops' to a new trap,
where it remains until it is again activated by thermal fluctuations.
Thus the state of the system can be described by the probability
$P(b,t)$ that it is in a trap of depth $b$ at time~$t$
(in the many--particle interpretation, $P(b,t)$ is the distribution
of barriers in all the subsystems).
Given that the barrier heights are distributed according to
some {\em prior} distribution $\rho(b)$,
then $P(b,t)$ evolves according to the master equation

\begin{equation}
\frac{1}{\act_{0}}\frac{\partial P(b,t)}{\partial t} =
\mbox{}-{{\e}^{-b/T}}\,P(b,t)+\act(t)\,\rho(b)\:,
\label{e:original_model}
\end{equation}

\noindent{}where $\act_{0}\,\act(t)$ is the total rate of hopping at time $t$,

\begin{equation}
\act_{0}\,\act(t) =
\act_{0}\,\int_{0}^{\infty}\,{{\e}^{-b/T}}\,P(b,t)\,{\rm db}\:.
\label{e:definition_of_rate}
\end{equation}

\noindent{}The first and second terms on the right hand side of
(\ref{e:original_model}) correspond to hopping out of and into
traps of depth~$b$, respectively.
For simplicity is has been assumed that $\rho(b)$ is independent
of $t$ and~$T$, and barrier heights before and after a hop are uncorrelated.

When a steady state solution of (\ref{e:original_model}) exists,
it takes the form

\begin{equation}
P_{\infty}(b)\equiv \lim_{t\rightarrow\infty}P(b,t)=
\act_{\infty}\,{\e}^{b/T}\,\rho(b)\:\:,
\label{e:orig_steady_state}
\end{equation}

\noindent{}using \mbox{$\act_{\infty}=\lim_{\,t\rightarrow\infty}\act(t)$}.
For a prior distribution with an exponential
tail $\rho(b)\sim{\e}^{-b/b_{0}}$ (which may be generic to glassy
systems~\cite{monthus_and_bouchaud}),
\mbox{$P_{\infty}(b)$} can only be normalised for \mbox{$T>b_{0}$}\,,
when it takes the form of a Boltzmann distribution.
No normalisable steady state solution exists for \mbox{$T\leq b_{0}$}\,,
when the system perpetually evolves into deeper and deeper traps.
In this manner the model exhibits a phase transition from equilibrium
to non-equilibrium behaviour at \mbox{$T=b_{0}$}\,,
{\em ie.} a {\em glass} transition.

\subsection{Master equation for strong interactions}
\label{s:new_model}

Although the basic trap model described above exhibits a
simple and elegant glass transition, it has the undesirable feature
that the mean barrier height diverges with time
when \mbox{$T\leq b_{0}$}\,.
This suggests that there may be some mechanism currently lacking
from the model which
eventually halts its progression into ever deeper traps.
For instance, it is known that the system will equilibrate at
very long times if it can only sample a finite number of
different configurations~\cite{bouchaud}.
Here we argue that introducing a suitable form of
interaction between the subsystems can achieve a similar effect.

Bouchaud {\em et al.} have considered the effect of allowing
each hop to slightly alter the barriers of
other subsystems~\cite{weakly}, so we are now working
strictly in the many-particle interpretation.
In their scheme, the barrier distribution $P(b,t)$ is perturbed
at a rate proportional to the overall hopping rate~$\act(t)$,
giving rise to an extra term in the master equation
(\ref{e:original_model}) of the form

\begin{equation}
\act(t)\int_{0}^{\infty}\mbox{\CLT}
(b'-b)\left( P(b',t)\rho(b)-P(b,t)\rho(b') \right)\,{\rm db'}\:.
\label{e:transfer}
\end{equation}

\noindent{}For weak interactions, \CLT$(b)$ is narrow and (\ref{e:transfer})
reduces to a diffusion--like term which was shown to not alter the essential
nature of the glass transition~\cite{weakly}.

Here we consider the opposite extreme of a broad \mbox{\CLT$(b)\equiv z$},
where $z$ is a constant.
This corresponds to a form of {\em strong} interaction in which
every hopping subsystem causes $z$ other
subsystems to have their barriers assigned {\em entirely} new values.
This erases the memory of the system at a rate $z\,\act(t)$
and, as shall be demonstrated below, ultimately removes the
glass transition.
Subsystems with very large barriers that would not normally
hop until very late times will now interact instead,
which on average lowers their barriers towards the mean
of $\rho(b)$ and allows them to hop
much earlier than they would in the absence of strong interactions.
Thus the interactions introduce a form of {\em dynamically
generated cut-off}\, into the system.

It is straightforward to integrate (\ref{e:transfer}) with
\mbox{\CLT$(b)=z$} to give the new master equation,

\begin{equation}
\frac{1}{\act_{0}}\frac{\partial P(b,t)}{\partial t} = 
\mbox{}-{{\e}^{-b/T}}\,P(b,t)
+\act(t)\,\rho(b)
+z\,\act(t)\Big(\rho(b)-P(b,t)\Big)\:,
\label{e:master_eqn}
\end{equation}

\noindent{}with the definition of $\act(t)$ unchanged from
(\ref{e:definition_of_rate}).
The steady state solution of (\ref{e:master_eqn}) is
non-Boltzmann,

\begin{eqnarray}
P_{\infty}(b)&=&\frac{z+1}{z}
\left(1+\frac{{\e}^{-b/T}}{\az}\right)^{-1}\rho(b)\:,
\label{e:steady_state}
\end{eqnarray}

\noindent{}so $P_{\infty}(b)\sim\rho(b)$ as \mbox{$b\rightarrow\infty$}
for all $T$, {\em  ie.} $P_{\infty}(b)$ is always normalisable
when \mbox{$z>0$}.
This suggests that there is no true finite temperature glass transition,
although it has yet to be shown that
the system always approaches this stationary solution.
Numerical evidence that this is indeed the case is presented in
Section \ref{s:lattice}, where it is also argued that glassy
behaviour may persist over intermediate timeframes at low temperatures.
For the remainder of this section we restrict our attention to analysis
of the stationary solution~(\ref{e:steady_state}).

\subsection{Steady state solution}

For the case of a simple exponential prior
\mbox{$\rho(b)=\frac{1}{b_{0}}{\e}^{-b/b_{0}}$}\,,
it is possible to derive an exact expression for $\act_{\infty}$
for \mbox{$z>0$}.
Details of this calculation are given in \ref{s:derivation_of_act};
here we just quote the final result.
In terms of the {\em reduced temperature}
\mbox{$x=T/b_{0}$}\,, the expression for $\act_{\infty}$ is

\begin{equation}
(\az)^{x}+\frac{\sin\pi x}{\pi x} \left\{
\frac{1}{z+1}+\,x\,\sum_{k=0}^{\infty}\frac{(-\az)^{k}}{k-x}
\right\}=0\:\:.
\label{e:act_solution}
\end{equation}

\noindent{}It may at first appear that the factor of \mbox{$\sin\pi x$}
on the left hand side of (\ref{e:act_solution})
would imply that $\act_{\infty}\rightarrow0$ for integer
values of~$x$.
However, there is also a simple pole in the sum for such~$x$,
and in fact these two effects cancel.
The equivalent expression to (\ref{e:act_solution}) for integer
$x$ is

\begin{equation}
\frac{(\az)^{-x}}{x(z+1)}=
(-1)^{x}\ln\left(1+\frac{1}{\az}\right) +
\sum_{k=0}^{x-1}
\frac{(-1)^{k}\,(\az)^{k-x}}{x-k}\:\:.
\label{e:act_integer}
\end{equation}

\begin{figure}
\centerline{\psfig{file=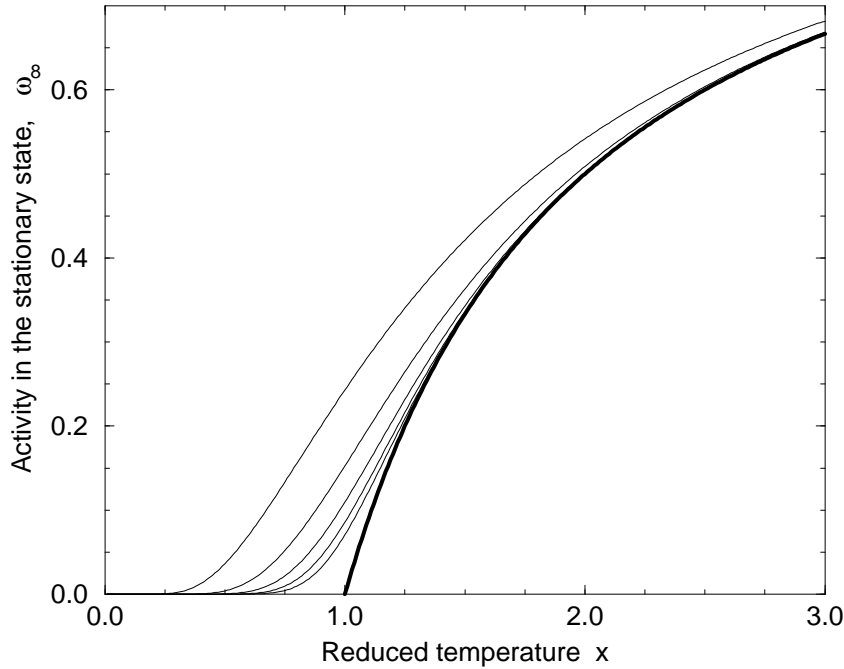,width=5in}}
\caption{Plot of $\act_{\infty}$ against the reduced temperature
\mbox{$x=T/b_{0}$}\,, found from (\ref{e:act_solution}) and
(\ref{e:act_integer}) by the method of interval bisection.
From top the bottom, the thin lines refer to \mbox{$z=10^{-1}$},
$10^{-2}$, $10^{-3}$, $10^{-4}$ and $10^{-5}$ respectively.
For comparison the \mbox{$z\equiv0$} solution
\mbox{$\act_{\infty}=1-1/x$} is plotted as a thick line.}
\label{f:act_vs_temp}
\end{figure}

\noindent{}The numerical plot of (\ref{e:act_solution}) and
(\ref{e:act_integer}) given in Fig.~\ref{f:act_vs_temp}
confirms the monotonic dependence of $\act_{\infty}$ on temperature.

One might expect that $P_{\infty}(b)$ for small $z$
could be expressed as just an \CLO$(z)$ perturbation around
the \mbox{$z\equiv0$} solution given
in~(\ref{e:orig_steady_state}).
In fact this is only true for temperatures \mbox{$x>2$}, as we now demonstrate.
As \mbox{$z\rightarrow0$}, (\ref{e:act_solution}) can be expanded in
powers of $z$ and the higher order terms dropped.
The identification of the leading order term depends upon the
value of~$x$.
For \mbox{$x>1$}, the leading order term is \CLO$(\az)$ and

\begin{equation}
\act_{\infty}=\left(1-\frac{1}{x}\right)
\left(\,1+\mbox{\CLO}(z)\,\right) \:\:,
\label{e:order_z_perturb}
\end{equation}

\noindent{}which is just an \CLO$(z)$ perturbation around the
original solution~(\ref{e:orig_steady_state}).
This is not true for \mbox{$x<1$}, when the
$(\az)^{x}$ term is now leading order and

\begin{equation}
\act_{\infty}\approx
\left(\frac{z}{z+1}\right)^{\frac{1}{x}-1}\left(
\frac{\sin\pi x}{\pi x}
\right)^{\frac{1}{x}}\:\:.
\label{e:act_lt_1}
\end{equation}

\noindent{}This holds as long as
the \CLO$(\az)^{x}$ term is much larger than the \CLO$(\az)$ term,
which corresponds to \mbox{$z^{1-x}\ll1$}.
This should be contrasted to the original model, where
$\act_{\infty}$ is not even defined for \mbox{$x\leq1$}.
Repeating this procedure for integer $x$
produces the same expression as (\ref{e:order_z_perturb}) for
\mbox{$x\geq2$} and \mbox{$\act_{\infty}\sim-(\ln z)^{-1}$} for \mbox{$x=1$}.

The consequence of having \mbox{$z>0$} becomes apparent at higher
temperatures if one considers the distribution of barriers as a whole,
rather than just~$\act_{\infty}$\,.
By substituting (\ref{e:order_z_perturb}) into the expression
for $P_{\infty}(b)$ given in (\ref{e:steady_state}) and expanding in powers
of~$z$, one readily sees that
$P_{\infty}(b)$ can only be expressed as an \CLO$(z)$ perturbation around
the original solution when \mbox{$x>2$};
for \mbox{$x\leq2$}, the assumption of a linear expansion breaks down.
This higher-temperature divergence can be explained by observing that,
when \mbox{$z\equiv0$},
the mean time between hops diverges in the stationary state
for \mbox{$x\leq2$}~\cite{monthus_and_bouchaud}.
By contrast, when \mbox{$z>0$} the mean time between interactions
approaches \mbox{$(\az)^{-1}$}, which is {\em always} finite.
Thus the interactions will always be significant when \mbox{$x\leq2$},
no matter how small $z$ may be.

\subsection{Solution at the critical point $T=0^{+}$}
\label{s:mf_baksnep}

An important consequence of introducing this new form of interaction
is that the system now becomes critical in the limit
\mbox{$T\rightarrow0^{+}$}.
This can be most clearly seen by considering a finite system
of $N$ subsystems with barriers $b_{i}$\,, \mbox{$i=1\ldots N$}.
For small $T$ the system will remain static for long periods,
but when a subsystem {\em does} eventually hop, the
probability that it had the barrier $b_{i}$ is

\begin{equation}
p_{i}=\frac{{\e}^{-b_{i}/T}}{\sum_{j=1}^{N}\,{\e}^{-b_{j}/T}}\:\:.
\label{e:partition}
\end{equation}

\noindent{}Since the system is finite it is always possible to
identify a unique minimum barrier $b_{\rm min}$
(assuming that $\rho(b)$ does not contain any delta-function peaks).
Suppose it is subsystem $j$ that has the barrier $b_{\rm min}$\,,
then inspection of (\ref{e:partition}) shows that
$p_{j}\rightarrow1$ as \mbox{$T\rightarrow0^{+}$}
while $p_{i}\rightarrow0$ for all \mbox{$i\neq j$}.
Thus the identification of the activated subsystem is entirely
deterministic -- it is {\em always} the one with the
smallest barrier.
This is now algorithmically identical to the Bak-Sneppen
model, already widely studied in the context
of self-organised criticality~\cite{BS,BSRev}.
Hence \mbox{$T=0^{+}$} is a critical point of the current model.
Note that the timescale in the Bak-Sneppen model is
normalised to precisely one hop per unit time,
whereas in our model the time between hops can vary.
However, as explained below, the maximum time between hops
is of the order of \mbox{${\e}^{b_{\rm c}/T}$} with $b_{\rm c}$ constant,
which is finite for all \mbox{$T>0$}.
Hence we do not anticipate any problems with {\em e.g.} diverging moments
of the distribution of hopping times.

The Bak-Sneppen model is usually defined on a lattice, in which
interactions only occur between nearest neighbours.
This suggests that the current model can also be given a lattice
interpretation, and this modification is considered in
section~\ref{s:lattice}.
However, a mean field solution to the Bak-Sneppen model already
exists~\cite{BS_MF1,BS_MF2}, which should be the same as the
\mbox{$T\rightarrow0^{+}$} solution of the current model.
This indeed proves to be the case for the stationary
distribution~$P_{\infty}(b)$, derived in \ref{s:T_0_solution}
for arbitrary $\rho(b)$,

\begin{eqnarray}
P_{\infty}(b)&\approx&\frac{z+1}{z}
\:\frac{\rho(b)}{1+{\e}^{(b_{\rm c}-b)/T}}
\hspace{1cm}\mbox{for small $T$,}
\label{e:bak_sneppen}\\
&\sim&\frac{z+1}{z}\:\theta(b-b_{\rm c})\:\rho(b)
\hspace{1cm}\mbox{as $T\rightarrow0^{+}$,}
\label{e:theta_sol}
\end{eqnarray}

\noindent{}where $\theta(x)$ is the usual theta function and
$b_{\rm c}$ is defined by

\begin{equation}
\int_{0}^{b_{\rm c}}\,\rho(b)\,{\rm db}=\frac{1}{z+1}\:\:.
\label{e:b_c_definition}
\end{equation}

\noindent{}This is also the mean field solution to the Bak-Sneppen model,
to within \CLO$(1/N)$.
Note that (\ref{e:bak_sneppen})
was found by taking \mbox{$T\rightarrow0^{+}$}
{\em after} taking the thermodynamic limit \mbox{$N\rightarrow\infty$}.
That it agrees with the mean field solution of the Bak-Sneppen model,
which corresponds to taking \mbox{$T\rightarrow0^{+}$}
{\em before} \mbox{$N\rightarrow\infty$},
suggests that the order of the limits may not be important
for time-independent quantities, although this may not hold
for time-dependent measures.

The discontinuity in $b$ in~(\ref{e:theta_sol}) arises from the existence
of a single characteristic time between interactions of the
order of \mbox{$\act_{0}\,{\e}^{b_{\rm c}/T}$}.
The expected time until a subsystem with a barrier $b$ hops
is $\act_{0}\,{\e}^{b/T}$, so subsystems with \mbox{$b\ll b_{c}$}
will almost certainly hop before interacting,
whereas those with \mbox{$b\gg b_{c}$} will almost certainly
interact before hopping.
Only barriers \mbox{$b\approx b_{c}$} will have comparable
probabilities of either hopping or interacting.
As $T$ is lowered the distribution of hopping times
becomes extremely broad, corresponding to an increasingly narrow
range of $b$ for which the rates of hopping to interaction are comparable,
eventually collapsing onto the single point $b_{c}$ at \mbox{$T=0^{+}$}.
Since hopping takes place over timescales much shorter than interactions,
there will only be a vanishingly small proportion of subsystems with
\mbox{$b<b_{\rm c}$} at any given time, as implied by the theta
function in~(\ref{e:theta_sol}).


This suggests the classification of subsystems as either
`active' (those with \mbox{$b<b_{\rm c}$}) or `inactive'
(those with \mbox{$b>b_{\rm c}$}).
In this heuristic scheme, active subsystems may become inactive by hopping
into a trap with a barrier greater than~$b_{c}$\,,
but inactive subsystems can only become active via interactions.
This description is somewhat reminiscent of the contact process,
a critical model which falls into the universality class of
directed percolation~\cite{CP}.
However, the contact process only becomes critical when its control
parameter is set to some finite value -- in effect,
its value of $b_{c}$ has to be set `by hand.'
By contrast, the current model automatically finds $b_{c}$ in
the limiting case $T\rightarrow0^{+}$.
It is in this sense that the Bak-Sneppen model
can be said to be ``self-organised'' critical.
A fuller discussion of the relationship between this model and
self-organised criticality will be made elsewhere~\cite{itshmeee}.
Note that the Bak-Sneppen model and the contact process have different
driving terms and fall into different universality classes~\footnote[1]{The
contact process also has slightly
different interaction terms to the Bak-Sneppen model, but these can
be altered without changing its universality
class~\cite{mod_int,grassberger}.}.

\section{Finite dimensional lattice model}
\label{s:lattice}

The analysis of the preceeding section is mean field in
the sense that the interactions have been assumed to act
homogeneously throughout the system.
In this section the model is given spatial definition
by identifying each subsystem with a
single site of a $d$--dimensional lattice.
The subsystems become thermally activated as before, but now
each activation event can only alter the barriers of adjacent
lattice sites, so the interactions are strictly short ranged.
Only a \oned lattice is considered here, but the model
can be defined in higher dimensions in a similar manner.

The lattice model is defined as follows.
A single barrier height $b_{i}$ is stored in each site
\mbox{$i=1\ldots N$} of a \oned lattice of size~$N$,
with initial conditions to be specified below.
At the start of each time step, every site is checked to see whether it
becomes activated, which it does with a probability~${\e}^{-b_{i}/T}$.
Any activated subsystems hop to a new trap
with a new barrier $b_{i}$ drawn from $\rho(b)$.
Furthermore, the barriers in the adjacent sites $b_{i-1}$ and $b_{i+1}$
are also assigned new values -- this represents the strong interactions.
Note that in these scheme each site hops at most once per time step.
This discrete time variable differs from the continuous time
employed in the mean field equations in section~\ref{s:mf_analysis},
but should make little difference when the overall rate
of activity is low.

Thus every activation event causes {\em three} barriers
to change value -- the activated site itself, and the barriers
in both of the adjacent sites.
This would appear to fix \mbox{$z=2$}.
However, smaller values of $z$ can be incorporated by stipulating that
$b_{i-1}$ and $b_{i+1}$ only change values with a probability $z/2$,
where now \mbox{$0\leq z\leq2$}.
Simulation results for various $z$ and
$T$ are discussed in sections
(\ref{s:activity}) to (\ref{s:bs_aging}) below.
Periodic boundary conditions have been assumed throughout,
so \mbox{$b_{N+1}\equiv b_{1}$}\,.
The prior distribution was taken to be the simple exponential
\mbox{$\rho(b)=\frac{1}{b_{0}}{\e}^{-b/b_{0}}$}.
Care was also taken to ensure that the supplied random number generator
gave good statistics deep in the tail of an exponential distribution.

\subsection{Existence of a stationary solution}
\label{s:activity}

The simulations were first employed to see whether or not there exists
a stationary solution in the lattice model, and if so,
how robust it is with respect to the initial conditions.
To test this, two very different initial configurations
were used.
The first was  an initial `quench' where all the $b_{i}$
were drawn from $\rho(b)$ but spatially uncorrelated.
These runs were then repeated with the same parameters but starting
with every $b_{i}$ arbitrarily large except for a single `seed'
at the origin with \mbox{$b_{0}=0$} (or any finite value).
In both cases the barrier distribution $P(b,t)$ appeared to converge
to the same distribution $P_{\infty}(b)$,
which remained steady within the timeframe of the simulations,
typically $10^{7}$ time steps
(although the situation is not so clear for the \mbox{$T=0^{+}$}
limit in low dimensions; see section~\ref{s:bs_aging}).
Since the initial conditions were so very different,
it seems likely that this same distribution
is approached irrespective of the initial state.
The stationarity of this solution
is confirmed in section~\ref{s:aging}.

\begin{figure}
\centerline{\psfig{file=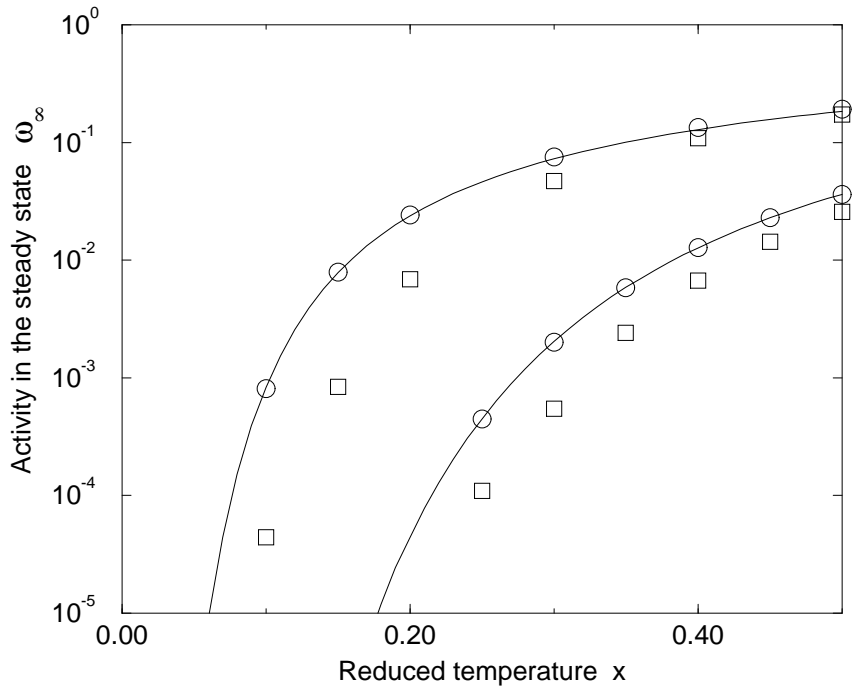,width=5in}}
\caption{Plot of $\act_{\infty}$ against $x$ for \mbox{$z=1$} (upper
data sets) and \mbox{$z=0.1$} (lower data sets).
In both cases, the circles and squares correspond to simulation results
for the random nearest neighbour and \oned models, respectively.
The system size was \mbox{$N=10^{4}$} and
the estimated error bars are smaller than the symbols.
The solid lines are the mean field predictions from
(\ref{e:act_solution}) and (\ref{e:act_integer}).}
\label{f:comp}
\end{figure}

Although it has already been shown that the mean field equations
of section~\ref{s:new_model} admit a stationary solution,
it is by no means clear that it is always, if ever, reached.
To investigate this, the simulations were also repeated using a `random nearest
neighbour' (RNN) algorithm in which the neighbours of the hopping
sites are chosen entirely at random from the remainder of the system,
with new neighbours being chosen at every time step.
Apart from finite size effects and the discrete time variable,
this should behave identically to the mean field model.
Simulations of the RNN model show that
the same stationary solution was reached for
both sets of initial conditions, as in the \oned case.
Furthermore the numerical estimate of $\act_{\infty}$ was found to
agree well with the
theoretical predictions (\ref{e:act_solution}) and (\ref{e:act_integer}),
as plotted in Fig.~\ref{f:comp}.

Also plotted in Fig.~\ref{f:comp} are the equivalent
values of $\act_{\infty}$ for the \oned model,
which consistently lie under the mean field values.
This is because the interactions are now spatially correlated with
the hopping sites and are more likely to occur in regions of high activity,
{\em ie.} low barriers.
Since the interactions tend to decrease barrier values towards the
mean of \mbox{$\rho(b)$}, their effect on $P(b,t)$
will be {\em weaker} in the \oned model than the corresponding
mean field system.
Thus the stationary state will not be reached until much later
times, when the activity will have decayed to smaller values,
as in Fig.~\ref{f:comp}.
In the \mbox{$T\rightarrow0^{+}$} limit
the limiting activity is \mbox{$\act_{\infty}\sim{\e}^{-b_{\rm c}/T}$}
(see \ref{s:T_0_solution}),
so as a corollary
$b_{\rm c}$ should be {\em higher} in lower dimensions,
and indeed this has already been observed in the
Bak-Sneppen model~\cite{BS_MF1,BS_MF2}.

\subsection{Interruption of aging}
\label{s:aging}

An important feature of many glassy systems is that their behaviour
depends upon the time since the initial quench, a property known
as `aging.'
This phenomenom can be quantified by use of the two--time correlation
function $C(t_{\rm w}+t,t_{\rm w})$,
defined here as the proportion of sites that have not hopped between
times $t_{\rm w}$ and~$t_{\rm w}+t$\, following a quench at \mbox{$t=0$}.
In Bouchaud's model the system becomes
{\em time--translationally invariant} above the glass transition,
{\em ie.} $C(t_{\rm w}+t,t_{\rm w})\rightarrow C_{\rm eq}(t)$
as \mbox{$t_{\rm w}\rightarrow\infty$} when \mbox{$x>1$}.
By contrast, when \mbox{$x<1$} the system behaviour
cannot be decoupled from $t_{\rm w}$ and instead one finds
$C(t_{\rm w}+t,t_{\rm w})\rightarrow C_{\rm neq}(t/t_{\rm w})$,
demonstrating that the system ages~\cite{monthus_and_bouchaud}.

\begin{figure}
\centerline{\psfig{file=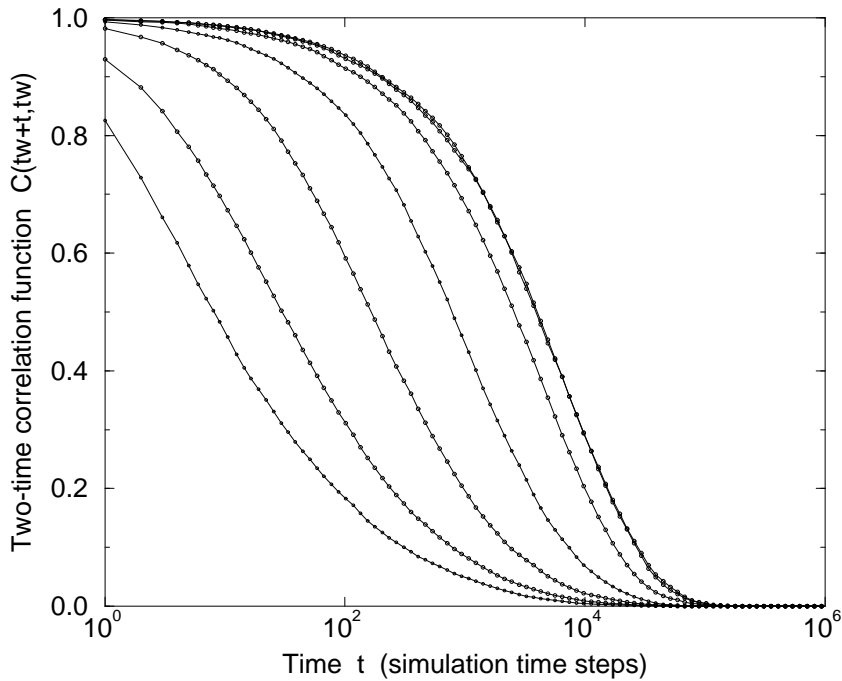,width=5in}}
\caption{The two--time correlation function $C(t_{\rm w}+t,t_{\rm w})$
plotted against $t$ for a \oned lattice with $N=10^{4}$,
\mbox{$z=0.1$}, \mbox{$x=0.35$} and \mbox{$\act_{0}=1$}.
From left to right, \mbox{$t_{\rm w}=10^{0}$},
$10^{1}$, $10^{2}$, $10^{3}$, $10^{4}$, $10^{5}$ and $10^{6}$.
Note that the last two lines overlap.
The system was started from an initial quench with
the $b_{i}$ drawn from $\rho(b)$ and uncorrelated.}
\label{f:twotime}
\end{figure}

To confirm that the simulations do indeed converge to a stationary
solution, it is necessary to show that $C(t_{\rm w}+t,t_{\rm w})$
converges to a time-translationally invariant solution~$C_{\infty}(t)$.
This has been checked in both \oned and RNN simulations,
and in all cases stationarity was reached after waiting times
\mbox{$t_{\rm w}\gg \tia$}\,,
where $\tia$ depends on $z$, $T$ and the dimensionality.
A typical example is given in Fig.~\ref{f:twotime}.
An estimate of $\tia$ for low temperatures
in the mean field can be found by following the
procedure described in \cite{monthus_and_bouchaud}.
This involves substituting the trial solution

\begin{equation}
P(b,t)={\textstyle \frac{1}{T}}\,u\,\phi(u)\:\:,
\end{equation}

\noindent{}expressed here in terms of the dimensionless variable
\mbox{$u=\frac{1}{\act_{0}t}\,{\e}^{b/T}$}.
For \mbox{$\rho(b)=\frac{1}{b_{0}}{\e}^{-b/b_{0}}$}
the master equation (\ref{e:master_eqn}) then becomes

\begin{equation}
u^{2}\frac{\rm d\phi(u)}{\rm du}+(u-1)\,\phi(u)=
\left(\int_{1/\act_{0}t}^{\infty}\frac{\phi(v)}{v}\,{\rm dv}\right)
\left(zu\,\phi(u)-\frac{x(1\!+\!z)}{(\act_{0}\,ut)^{x}}\right)\:\:,
\label{e:scaling_soln}
\end{equation}

\noindent{}where the integral is just $\omega(t)$
expressed in terms of $\phi$ and~$u$.
Dimensional analysis suggests that the first term on the right
hand side of (\ref{e:scaling_soln})
can be neglected for small~$t$.
This allows a $t$-independent scaling solution to
be found which exhibits aging~\cite{monthus_and_bouchaud}.
However, if \mbox{$z>0$} then, as \mbox{$t\rightarrow\infty$},
the second term will become negligible
and it is straightforward to show that no
self-consistent scaling solution exists.
From this we infer that the aging is {\em interrupted} at some finite
time.
The crossover time $\tia$ corresponds to times when both terms
are of comparable magnitude, which can be estimated by
dimensional analysis to be

\begin{equation}
\tia\sim \left(\frac{1+z}{z}\right)^{1/x}\sim\:{\e}^{b_{c}/T}\:\:,
\label{e:tia}
\end{equation}

\noindent{}where $b_{\rm c}$ is the same as that defined in
(\ref{e:b_c_definition}).
Note that $\tia$
diverges rapidly as \mbox{$T\rightarrow0$}, so aging behaviour
may persist over intermediate (possibly experimental) timeframes
at low temperatures.

\subsection{Aging in the $T=0^{+}$ limit}
\label{s:bs_aging}

The analysis of the \mbox{$T=0^{+}$} limit presented in
section~\ref{s:mf_baksnep} was limited to the stationary
barrier distribution~$P_{\infty}(b)$.
We now turn to consider time-dependent behaviour, which
requires careful consideration of the choice of timescale.
Clearly as \mbox{$T\rightarrow0^{+}$} the rate of hopping in a finite system
becomes vanishingly small and so the timescale must
be normalised in some manner to attain non-trivial behaviour.
Two different timescales will be considered here.
The first is the {\em event timescale} $\tev$ in which
precisely one subsystem from a total of $N$ hops per
unit~$\tev$.
This is the usual choice in the Bak-Sneppen model.
The second time variable we shall consider is the
{\em avalanche timescale}~$\tav$,
defined such that $\tav$ increases by one unit for every avalanche
in the system, where an avalanche is defined below.
Although not usually considered in connection with the Bak-Sneppen model,
$\tav$ is more in line with the timescale employed in
{\em e.g.} the sandpile model~\cite{dhar_rev}.
The existence of aging behaviour depends crucially
upon which timescale is adopted,
as we now demonstrate.

\begin{figure}
\centerline{\psfig{file=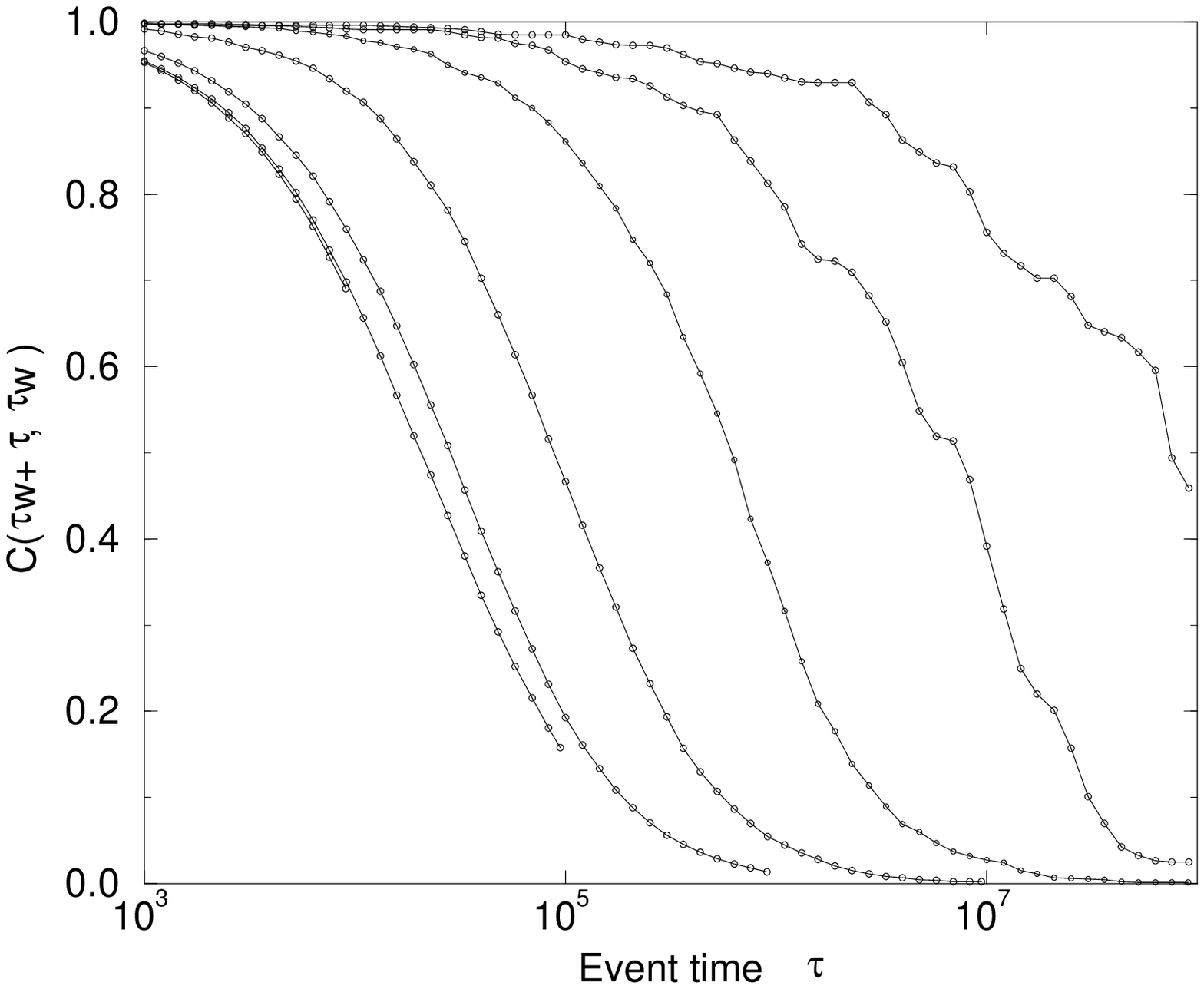,width=5in}}
\caption{$C(\tev_{\rm w}+\tev,\tev_{\rm w})$ versus $\tev$
for a \oned lattice of \mbox{$N=2\times10^{4}$} sites
in the \mbox{$T\rightarrow0^{+}$} limit in
event time $\tev$, making this identical to the Bak-Sneppen model.
From left to right, the different lines refer to
$\tev_{\rm w}=10^{0}$, $10^{1}$, $10^{2}$, $10^{3}$, $10^{4}$,
$10^{5}$, $10^{6}$, $10^{7}$ and $10^{8}$, respectively,
although the first 4 lines overlap.
The apparent cut-off for small $\tev_{\rm w}$ is purely
the result of an optimisation procedure employed to improve run times.}
\label{f:bs_aging}
\end{figure}

\begin{figure}
\centerline{\psfig{file=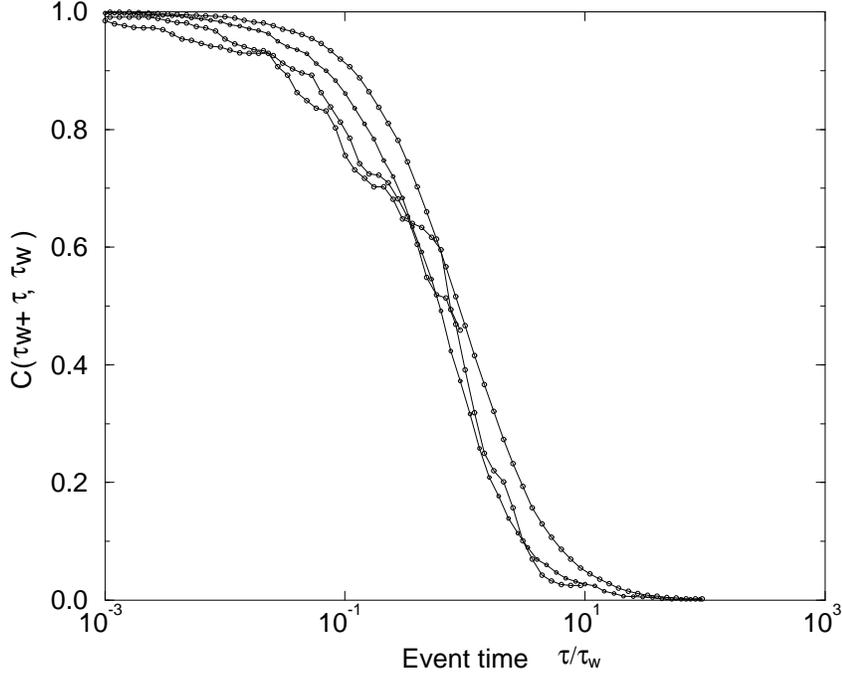,width=5in}}
\caption{The same data as in Fig.~\ref{f:bs_aging}
for \mbox{$\tau_{\rm w}\geq10^{5}$}, this time
plotted against \mbox{$\tau/\tau_{\rm w}$}\,, indicating
partial collapse.
From right to left (as viewed from the upper portion of
the graph), the lines refer to
\mbox{$\tau_{\rm w}=10^{5}$}, $10^{6}$, $10^{7}$ and $10^{8}$
respectively.}
\label{f:bs_aging2}
\end{figure}

Boettcher {\em et al.} have found that the Bak-Sneppen model
exhibits a form of aging in one and two dimensional simulations starting
from a single active seed~\cite{age_bs_1,age_bs_2}.
We have adapted our code to employ event time
in simulations starting from an initial quench
and have also observed aging in the \oned model,
but {\em not} in the RNN model.
Results from the \oned simulations are given in Fig.~\ref{f:bs_aging},
which appears to exhibit aging for the longest times we
were able to simulate, up to \mbox{$\tev_{\rm w}=10^{8}$}
on a \mbox{$N=2\times10^{4}$} lattice, although the statistics become
very noisy at later times.
For large \mbox{$\tev_{\rm w}$} the data shows some indications of
collapse onto a scaling function $C_{\infty}(\tev/\tev_{\rm w})$,
as demonstrated in Fig.~\ref{f:bs_aging2}.
By contrast the results from the RNN simulations (not given)
fail to exhibit any form of aging whatsoever.

It is possible to provide a rough derivation of the aging behaviour
observed in our simulations by considering
the pattern of activity of hopping sites as the system evolves
from its initial quench.
This argument employs known results for the Bak-Sneppen model, which
we do not attempt to justify here; the reader is instead
referred to \cite{BSRev} for a complete description.
Initially the location of the hopping site moves around the system
at random, but as $\tev$ increases it tends to repeatedly visit
one part of the system for a finite time before jumping to
another, uncorrelated region.
These localised spatiotemporal regions of activity are known
as {\em avalanches}.
The expected duration of the avalanche
$\langle S\rangle$ and the expected number of sites covered
during the avalanche $\langle n_{\rm cov}\rangle$ diverge with
time according to~\cite{BSRev}

\begin{eqnarray}
\langle S\rangle&\sim&\left(\frac{\tev}{N}\right)^{\frac{\gamma}
{\gamma-1}}\:,
\label{e:scal1}\\
\langle n_{\rm cov}\rangle &\sim&
\left(\frac{\tev}{N}\right)^{\frac{1}{\gamma-1}}\:\:,
\label{e:scal2}
\end{eqnarray}

\noindent{}where the critical exponent $\gamma$
depends on the dimensionality.
In low dimensions \mbox{$\gamma>1$}, but \mbox{$\gamma=1$} in
high dimensions where the scaling relations
given by (\ref{e:scal1}) and (\ref{e:scal2}) break down.
In this case the active site
never remains localised in any one part of the system
for any significant time and the following argument does not apply.

As before, let $C(\tev_{\rm w}+\tev,\tev_{\rm w})$ be the proportion of the
system that has not hopped between times $\tev_{\rm w}$ and
$\tev_{\rm w}+\tev$.
Clearly $C$ will decrease when the avalanche jumps to a new
part of the system.
If we now coarse grain over length and time scales much larger
than $\langle S\rangle$ and $\langle n_{\rm cov}\rangle$,
then, as consecutive avalanches are spatially uncorrelated,
we can write the following continuous differential equation for $C$,

\begin{equation}
\frac{\rm dC}{\rm d\tev}\approx -\frac{C}{N}
\frac{\langle n_{\rm cov}\rangle}
{\langle S\rangle}\:\:.
\label{e:dcbydt}
\end{equation}

\noindent{}$C$ only decays if the avalanche jumps to a part of the system
it has not already visited,
hence the factor of $C$ on the right hand side of (\ref{e:dcbydt}).
Furthermore each avalanche changes a proportion
\mbox{$\langle n_{\rm cov}\rangle/N$}
of the system and lasts for a time~$\langle S\rangle$, which accounts
for the remaining factors.
Integrating (\ref{e:dcbydt})
from $\tev_{\rm w}$ to $\tev_{\rm w}+\tev$ with
$C(\tev_{\rm w},\tev_{\rm w})=1$ gives

\begin{equation}
C(\tev_{\rm w}+\tev,\tev_{\rm w})\sim
\left(1+\frac{\tev}{\tev_{\rm w}}\right)^{-\alpha}
\end{equation}

\noindent{}where
\mbox{$\alpha=\lim_{\,\tev\rightarrow\infty}(\tev\,\langle n_{\rm cov}\rangle)/(N\,\langle S\rangle)$}.
That $C(\tev_{\rm w}+\tev,\tev_{\rm w})$ depends on $\tev$ and $\tev_{\rm w}$
only through the ratio $\tev/\tev_{\rm w}$ demonstrates that the system ages.
Although this derivation is sufficient to explain our simulation results,
it is not clear that the coarse graining assumption is valid
as $\tev\rightarrow\infty$, even for arbitrarily large systems.
It is also not obvious if this procedure can be extended to
account for the results of Boettcher {\em et al.}~\cite{age_bs_1,age_bs_2}.

If indeed the Bak-Sneppen model {\em does} age in event time~$\tev$,
this calls into question the usual assumption of stationarity
in most previous studies of the Bak-Sneppen model, since
an aging system does not obey time-translational invariance
and hence cannot be stationary.
It is possible that previous studies have relied
too heavily on one-time functions such as critical exponents~{\em etc.},
which may appear to tend to stationary values even when
two-time functions are still evolving.
This is a subtle question and further investigation would be desirable.
Note that the aging we have observed (apparent or otherwise)
is {\em dimensional--dependent} --- it {\em only} holds in low dimensions,
not in the mean field.
Hence the assumption of stationarity in section~\ref{s:mf_baksnep} is
still valid.

Turning now to consider avalanche time~$\tav$,
the same coarse-graining assumptions lead to the following
differential equation for $C(\tav_{\rm w}+\tav,\tav_{\rm w})$,

\begin{equation}
\frac{\rm dC}{\rm d\tav}\approx -\,
\frac{C\,\langle n_{\rm cov}\rangle}{N}\:\:,
\label{e:dc_avtime}
\end{equation}

\noindent{}which differs from (\ref{e:dcbydt}) by a factor
of~$\langle S\rangle$. This can be solved as before to give

\begin{equation}
C(\tav_{\rm w}+\tav,\tav_{\rm w})\sim
\exp\left\{
-\,a\left(\frac{\tav_{\rm w}}{N}\right)^{\frac{\gamma}{\gamma-1}}
\left[
\left(1+\frac{\tav}{\tav_{\rm w}}\right)^{\frac{\gamma}{\gamma-1}}-1
\right]
\right\}\:\:,
\label{e:noage}
\end{equation}

\begin{figure}
\centerline{\psfig{file=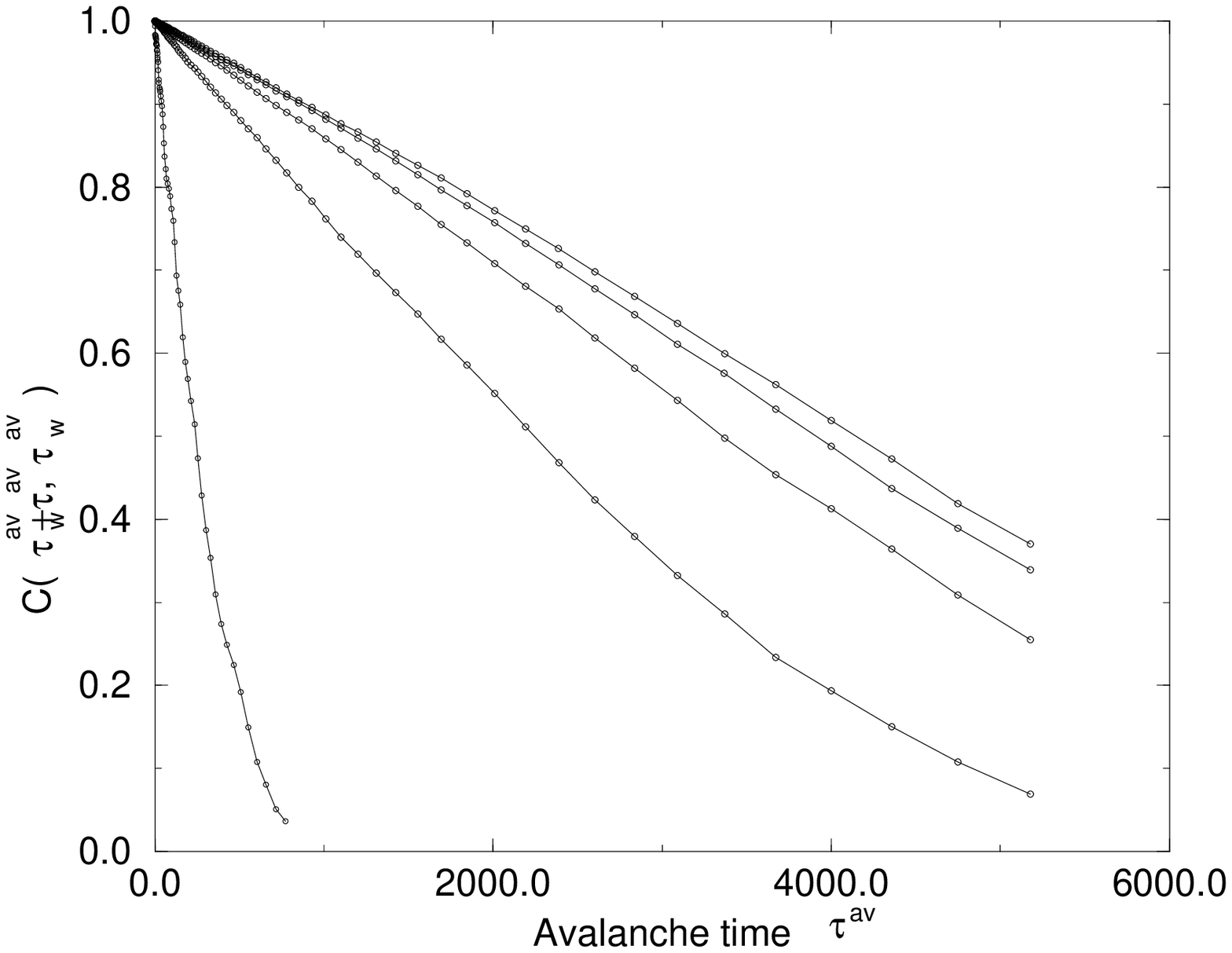,width=5in}}
\caption{$C(\tav_{\rm w}+\tav,\tav_{\rm w})$ measured in avalanche time
$\tav$ for a \oned lattice with \mbox{$N=10^{4}$} sites
in the \mbox{$T=0^{+}$} limit.
From upper-right to lower-left, the lines refer to
\mbox{$\tav_{\rm w}=R^{6}$}, $R^{7}$, $R^{8}$, $R^{9}$ and $R^{10}$
respectively, where \mbox{$R=2.512$}.}
\label{f:avtime}
\end{figure}

\noindent{}where $a$ is an arbitrary constant.
Since this expression tends to zero as \mbox{$\tav_{\rm w}\rightarrow\infty$}
for all $\tav>0$, we conclude that
the system does {\em not} age in avalanche time.
This may relate to the failure to observe aging in the sandpile
model~\cite{age_bs_1},
where a timescale similar to $\tav$ is usually employed.
Note that this conclusion also holds in the mean field case \mbox{$\gamma=1$},
although the form of~(\ref{e:noage}) is different.
The numerical plot of $C(\tav_{\rm w}+\tav,\tav_{\rm w})$ for
a \oned system is given in Fig.~\ref{f:avtime}.

\section{Conclusions}
\label{s:conc}

We have introduced the strongly-interacting trap model,
a version of Bouchaud's trap model which exhibits glassy behaviour
for low temperatures and intermediate timeframes, but ultimately
reaches stationarity as $t\rightarrow\infty$.
In the limit of zero temperature the model becomes critical in
all dimensions, including the mean field, so it has no
`lower critical dimension.'

A lattice version of the model was also introduced which appears to
behave qualitatively similar to the mean field model in all the
cases we have looked at, except one.
This is the apparent aging in the zero--temperature limit
in low dimensions in what we have termed `event time,'
which is absent in the mean field.
This would represent a form of dimensional--dependent aging,
but is currently only supported by numerical evidence
and hence remains somewhat speculative.
It is also not clear at what dimension the aging behaviour might start
to break down.

Although this work was originally motivated by the range
of spin glass models that have a critical point at \mbox{$T=0$},
we do not claim any precise relationship between these systems and our model.
We merely propose that our model may serve as a caricature of
glass models with zero-temperature criticality, which, by its
very simplicity, should allow for fuller analysis
of this class of systems,
both theoretically and numerically.
It may also serve as a link between glass theory and models
of self-organised criticality.

\section*{Acknowledgements}

The author gratefully acknowledges many fruitful discussions with
Mike Cates, Peter Sollich, Martin Evans, Mike Evans and Suzanne Fielding.
This work was funded under EPSRC grant no. GR/M09674.


\appendix

\section{Derivation of (\ref{e:act_solution}) and (\ref{e:act_integer})}
\label{s:derivation_of_act}

In this appendix we derive the expressions for $\act_{\infty}$
quoted in (\ref{e:act_solution}) and (\ref{e:act_integer}) for
non-integer and integer values of~$x=T/b_{0}$\,, respectively.
The steady state solution is found by substituting the form of
$P_{\infty}(b)$ given by (\ref{e:steady_state}) into the
definition of $\act(t)$ given in (\ref{e:definition_of_rate}).
This results in the following integral equation for $\act_{\infty}$\,,

\begin{equation}
\frac{1}{z+1}=\int_{0}^{\infty}\,
\frac{\rho(b)}{\az\,{\e}^{b/T}+1}
\,{\rm db}\:.
\label{e:starting_point}
\end{equation}

\noindent{}Substituting \mbox{$\rho(b)=\frac{1}{b_{0}}{\e}^{-b/b_{0}}$}
and making the change of variables \mbox{$u={\e}^{-b/T}$} gives

\begin{eqnarray}
\frac{1}{z+1}&=&x\,\int_{0}^{1}\,\frac{u^{x}}{u+\az}\,{\rm du}
\label{e:integral}\\
&=&\frac{x}{\az(1+x)}\,
F\Big(1,1+x\,;\,2+x\,;\,-(\az)^{-1}\Big)\:\:.
\label{e:hypergeometric}
\end{eqnarray}

\noindent{}For \mbox{$\az<1$} and non-integer $x$
the hypergeometric function $F$
can be rewritten as~\cite{prudnikov}

\begin{eqnarray}
\lefteqn{(\az)^{-1}\,F\Big(1,1+x\,;\,2+x\,;\,-(\az)^{-1}\Big)=}\hspace{1.5cm}\nonumber\\
&&\Gamma(2+x)\Gamma(-x)(\az)^{x}
-x\,\frac{\Gamma(2+x)\Gamma(x)}{(\,\Gamma(1+x)\,)^{2}}
\sum_{k=0}^{\infty}\frac{(-\az)^{k}}{k-x}\:\:,
\label{e:hypo_hyper}
\end{eqnarray}

\noindent{}where $\Gamma(x)$ is the usual gamma function.
Inserting this into (\ref{e:hypergeometric}) gives the final
expression~(\ref{e:act_solution}), where use has been made of the identity

\begin{equation}
\Gamma(x)\,\Gamma(-x)=-\,\frac{\pi}{x\sin\pi x}\:\:.
\end{equation}

\noindent{}The sum on the right hand side
of (\ref{e:hypo_hyper}) is ill defined when $x$ is an integer.
In this case, an
equivalent expression can be found, either by integrating (\ref{e:integral})
directly or by substituting \mbox{$x=n\pm\varepsilon$} into
(\ref{e:act_solution}) with $n$ an integer,
and letting \mbox{$\varepsilon\rightarrow0$}.
Both methods yield the same answer, given in (\ref{e:act_integer}).

\section{Derivation of the $T=0^{+}$ solution (\ref{e:bak_sneppen}) }
\label{s:T_0_solution}

It is possible to derive, in a non-rigorous manner,
explicit expressions for $\act_{\infty}$
and $P_{\infty}(b)$ for arbitrary $\rho(b)$ in the limit
\mbox{$T\rightarrow0^{+}$}.
The starting point is the same integral equation as in
(\ref{e:starting_point}),

\begin{equation}
\frac{1}{z+1}=\int_{0}^{\infty}\,
\frac{\rho(b)}{{\e}^{b/T+\ln(\az)}+1}
\,{\rm db}\:\:.
\label{e:T_0_start}
\end{equation}

\noindent{}This imposes significant constraints on the allowed
forms of~$\act_{\infty}$\,.
For instance, if \mbox{$T \ln(\az)\rightarrow0$} as \mbox{$T\rightarrow0$},
then (\ref{e:T_0_start}) would imply the contradictory result
\mbox{$z=\infty$}.
Indeed, if one assumes that \mbox{$\ln(\az)\sim T^{\alpha}$},
then self-consistency of (\ref{e:T_0_start}) demands that
\mbox{$\alpha=-1$} and therefore

\begin{equation}
\act_{\infty}\sim\frac{1}{z}\:{\e}^{-b_{\rm c}/T}\:\:,
\label{e:T_0_act_infty}
\end{equation}

\noindent{}where $b_{\rm c}$ is some arbitrary constant.

For $\ln(\az)=-b_{\rm c}/T$,  the denominator inside the integral
of (\ref{e:T_0_start}) is essentially constant except around the
region \mbox{$b\approx b_{\rm c}$}\,.
Since $\rho(b)$ is independent of~$T$, we can always choose $T$ small
enough so that $\rho(b)$ is slowly varying over this region,
assuming that $\rho(b)$ is continuous at~$b_{\rm c}$\,.
Hence the integral simplifies to

\begin{equation}
\frac{1}{z+1}=\int_{0}^{b_{\rm c}}\:\rho(b)\:{\rm db}\:\:,
\label{e:T_0_b_c}
\end{equation}

\noindent{}which fixes $b_{\rm c}$\,.
Substituting (\ref{e:T_0_act_infty})
into the general expression for $P_{\infty}(b)$ given in
(\ref{e:steady_state}) results in the final
expression (\ref{e:bak_sneppen}).
To give some idea of typical values of $b_{\rm c}$\,,
a simple exponential prior \mbox{$\rho(b)=\frac{1}{b_{0}}{\e}^{-b/b_{0}}$}
gives \mbox{$b_{\rm c}=b_{0}\ln(1+z^{-1})$}, whereas
\mbox{$b_{\rm c}=(1+z)^{-1}$} when $\rho(b)$ is uniformly distributed
on $[0,1]$.


\end{document}